# OMEGA FROM THE SKEWNESS OF
# THE COSMIC VELOCITY DIVERGENCE

by

F. Bernardeau[1,2], R. Juszkiewicz[3,4],
A. Dekel[4,5], F.R. Bouchet[4]

[1] CITA, University of Toronto, 60 St George St. Toronto, Ontario M5S 1A1, Canada
[2] Service de Physique Théorique, CE Saclay, 91191 Gif sur Yvette, France
[3] Copernicus Astronomical Center, ul. Bartycka 18, Warszawa, Poland
[4] Institut d'Astrophysique de Paris, CNRS, 98 bis Boulevard Arago F-75014, Paris, France
[5] Racah Institute of Physics, The Hebrew University, Jerusalem 91904, Israel

## ABSTRACT

We propose a method for measuring the cosmological density parameter $\Omega$ from the statistics of the expansion scalar, $\theta \equiv H^{-1}\nabla\cdot\boldsymbol{v}$, – the divergence of peculiar velocity, expressed in units of the Hubble constant, $H \equiv 100h$ km s$^{-1}$Mpc$^{-1}$. The velocity field is spatially smoothed over $\sim 10h^{-1}$Mpc to remove strongly nonlinear effects. Assuming weakly-nonlinear gravitational evolution from Gaussian initial fluctuations, and using second-order perturbative analysis, we show that $\langle\theta^3\rangle \propto -\Omega^{-0.6}\langle\theta^2\rangle^2$. The constant of proportionality depends on the smoothing window. For a top-hat of radius $R$ and volume-weighted smoothing, this constant is $26/7 - \gamma$, where $\gamma = -\mathrm{d}\log\langle\theta^2\rangle/\mathrm{d}\log R$. If the power spectrum is a power law, $P(k) \propto k^n$, then $\gamma = 3+n$. A Gaussian window yields similar results. The resulting method for measuring $\Omega$ is independent of any assumed biasing relation between galaxies and mass. The method has been successfully tested with numerical simulations. A preliminary application to real data, provided by the POTENT recovery procedure from observed velocities favors $\Omega \sim 1$. However, because of an uncertain sampling error, this result should be treated as an assessment of the feasibility of our method rather than a definitive measurement of $\Omega$.

## 1. INTRODUCTION

The standard model of large-scale structure assumes that the structure we observe today developed by gravitational instability from small-amplitude, random initial fluctuations, induced in the early universe, perhaps during an inflation phase (see Peebles 1993, for a review). The most natural hypothesis (see Bardeen et al. 1986 and references therein) is that the initial probability distribution of the density contrast, $\delta(\boldsymbol{x}) \equiv \rho(\boldsymbol{x})/\langle \rho \rangle - 1$, was Gaussian (here $\rho$ is the matter density, $\boldsymbol{x}$ are the comoving coordinates, and $\langle \ldots \rangle$ denote ensemble averaging). Under this hypothesis, the initial distribution has zero skewness, $\langle \delta^3 \rangle = 0$. With the growth of gravitational clustering, deviations from the initial Gaussian distribution develop. In particular, to second order in perturbation theory, the gravitationally induced skewness and variance are related by the scaling

$$S_3 \equiv \frac{\langle \delta^3 \rangle}{\langle \delta^2 \rangle^2} = \frac{34}{7} \; ; \tag{1}$$

the above result assumes $\langle \delta^2 \rangle \ll 1$ and $\Omega = 1$ (Peebles 1980, §18). Bouchet et al. (1992) derived $S_3$ for arbitrary $\Omega$. The $\Omega$-dependence turned out to be extremely weak, in agreement with earlier suggestions based on numerical integration of the equations of motion (Martel & Freudling 1991). To mimic the observational techniques of estimating the moments of $\delta$, Juszkiewicz et al. (1991;1993a) derived the $S_3$ parameter for a spatially smoothed field. They generalized earlier results (Grinstein & Wise 1986, Goroff et al. 1986), and showed that the filtering introduces a dependence of $S_3$ on the slope of the power spectrum of the density fluctuations, while the weakness of the $\Omega$ dependence of $S_3$ is preserved. These perturbative results were successfully tested against N-body simulations (Weinberg & Cole 1992; White 1992; Bouchet & Hernquist 1992; Juszkiewicz et al. 1993a,b; Lahav et al. 1993; Lucchin et al. 1994, Bernardeau 1993).

The weak dependence on a such a poorly known parameter, as $\Omega$, makes observational estimates of $S_3$ useful in constraining the initial probability distribution of $\delta$ and its *intrinsic* (as opposed to the gravitationally induced) skewness (Silk & Juszkiewicz 1991; Jaffe 1993, Bartlett et al. 1994). The observed density field of galaxies, under a reasonable assumption of local biasing relation between the galaxy density and the underlying mass density, was indeed found to be consistent with the scaling relation (1) (Coles & Frenk 1991; Bouchet et al. 1991;1993 – for the IRAS survey; Gaztañaga 1994 for the APM survey; for the effect of local biasing on $S_3$, see Fry & Gaztañaga 1993 and Juszkiewicz et al. 1993b). These results are consistent with the Gaussian initial conditions hypothesis and imply that the skewness present in the data seems to be entirely induced by gravity. Similar conclusions were reached by Nusser et al. (1993;94), who used the Zel'dovich (1970) approximation to reconstruct the initial probability distribution of $\delta$ from the IRAS data, and by Feldman et al. (1994), who investigated the spread of fluctuations in the measurements of the IRAS power spectrum.

The independence of $S_3$ on $\Omega$, so useful for the above purposes, makes it useless for measuring $\Omega$. However, as we will show in this paper, there is a close relative of $S_3$, which can be used to measure the cosmic mass density parameter. This is the skewness of the distribution of the divergence of the peculiar velocity (see Dekel 1994 for a review on cosmic flows). Radial peculiar velocities of more than 3000 galaxies have been measured so far within a sphere of radius $\sim 7000$ km s$^{-1}$ about the Local group, and the POTENT reconstruction method enables the processing of this data into a three-dimensional velocity field, smoothed at $\sim 12$ h$^{-1}$Mpc, based on the asserted potential-flow nature of quasilinear gravitating flows. This irrotational velocity field, $\boldsymbol{v}(\boldsymbol{x})$, can be characterized, up to an additive bulk velocity, by a single scalar field such as the velocity divergence

$$\theta(\boldsymbol{x}) \equiv \frac{1}{H}\boldsymbol{\nabla}\cdot\boldsymbol{v}(\boldsymbol{x}) , \qquad (2)$$

called the *expansion scalar* (see, *e.g.* Peebles 1980). Division by the Hubble constant, $H$, makes $\theta$ dimensionless.

The gravitational instability equations governing the evolution of fluctuations of a pressureless gravitating fluid in a standard cosmological background during the matter era are the continuity equation, the Euler equation of motion, and the Poisson field equation (*e.g.* Peebles 1993):

$$\begin{aligned}\dot{\delta} + \boldsymbol{\nabla}\cdot\boldsymbol{v} + \boldsymbol{\nabla}\cdot(\boldsymbol{v}\delta) &= 0 , \\ \dot{\boldsymbol{v}} + 2H\boldsymbol{v} + (\boldsymbol{v}\cdot\boldsymbol{\nabla})\boldsymbol{v} &= -\boldsymbol{\nabla}\Phi , \\ \boldsymbol{\nabla}^2\Phi &= (3/2)H^2\Omega\,\delta .\end{aligned} \qquad (3)$$

Here $\boldsymbol{x}, \boldsymbol{v}$, and $\Phi$ are the position, peculiar velocity and peculiar gravitational potential in comoving distance units, corresponding to $a\boldsymbol{x}$, $a\boldsymbol{v}$, and $a^2\Phi$ in physical units, with $a(t)$ the universal expansion factor, $t$ is the cosmological time and dots represent time derivatives. To *linear* order in $\delta$, $\boldsymbol{v}$ and $\Phi$, the growing-mode solution for the density fluctuation has a universal time dependence,

$$\delta_1(\boldsymbol{x},t) = D(t)\epsilon(\boldsymbol{x}) . \qquad (4)$$

The linear velocity is irrotational, derived from a velocity potential,

$$\boldsymbol{v}_1(\boldsymbol{x},t) = -\dot{D}(t)\boldsymbol{\nabla}\varphi(\boldsymbol{x}) , \qquad (5)$$

where $\varphi$ is time-independent and obeying the linearized Poisson equation $\boldsymbol{\nabla}^2\varphi = \epsilon$, *i.e.* the corresponding linear gravitational potential is $\Phi_1 = (3/2)H\Omega D\varphi$. The relation between the linear solutions of velocity and density in the linear approximation is then

$$\theta_1(\boldsymbol{x}) = -f(\Omega)\delta_1(\boldsymbol{x}) , \qquad (6)$$

where $f(\Omega) \equiv H^{-1}\dot{D}/D \approx \Omega^{0.6}$ (see Peebles 1993, eq. 5.120). This type of relation has been used in many different ways to determine $\Omega$ from the comparison of observed velocity and density fields (see Dekel 1994 §8 for a review), but since the density field is derived from the distribution of galaxies, such determinations are necessarily subject to an arbitrary choice of biasing relation between the densities of galaxies and mass.

In the present paper we derive $\Omega$ from the velocity field alone: our method is *independent* of galaxy biasing. The linear relation (6) indicates that a relation between the moments of $\theta$ analogous to equation (1) should depend strongly on $\Omega$ and may therefore be useful for measuring $\Omega$. Equation (6) also indicates that the perturbative technique used to compute the moments of $\delta$ should be applicable in the $\theta$ case as well. This calculation is presented in §2 for a general value of $\Omega$, taking into account smoothing with a Gaussian or a top-hat window function and an assumed power spectrum. The results, derived by perturbative calculations with respect to the initial fluctuations, can only be applied to a heavily-smoothed velocity field, where the fluctuation amplitudes are not too nonlinear. The validity of the perturbative approximation for deriving $\Omega$ under realistic nonlinear conditions is confirmed using numerical simulations in §3, where we also present preliminary results based on real observational data. Our method is discussed in §4.

## 2. MOMENTS OF THE EXPANSION SCALAR

### 2.1. Unsmoothed Fields

We wish to compute the relation between the skewness and variance of $\theta$, in analogy to equation (1), under gravitational evolution from Gaussian initial fluctuations. The perturbative expansion for $\theta$ is

$$\theta = \theta_1 + \theta_2 + O(\delta_1^3) , \qquad (7)$$

where $\theta_1$ is of order $\delta_1$, and $\theta_2 = O(\delta_1^2)$. The variance is then $\langle \theta^2 \rangle = \langle \theta_1^2 \rangle + O(\delta_1^3)$, so the leading term is simply

$$\langle \theta^2 \rangle = f(\Omega)^2 \langle \delta_1^2 \rangle . \qquad (8)$$

The third moment, by equation (7), is

$$\langle \theta^3 \rangle = \langle \theta_1^3 \rangle + 3\langle \theta_1^2 \theta_2 \rangle + O(\delta_1^5) . \qquad (9)$$

The skewness of $\theta_1$ vanishes as $\theta_1$ remains a Gaussian variable at all times, because it has a universal time dependence like $\delta_1$ (eqs. 4, 6). This leaves $3\langle \theta_1^2 \theta_2 \rangle$ as the leading term in equation (9), which implies that the perturbative expansion of $\theta$ must be carried out at least to second order.

In calculating $\theta_2$ we are guided by the analogous analysis of the density field (*e.g.* Bouchet et al. 1992; following Peebles 1980 §18), which yielded for the second-order term

$$\delta_2(\boldsymbol{x},t) = D(t)^2 \left( \frac{1}{2}\left[1 + \frac{F(t)}{D(t)^2}\right]\epsilon^2(\boldsymbol{x}) + \boldsymbol{\nabla}\epsilon(\boldsymbol{x})\cdot\boldsymbol{\nabla}\varphi(\boldsymbol{x}) + \frac{1}{2}\left[1 - \frac{F(t)}{D(t)^2}\right]\varphi_{,\alpha\beta}(\boldsymbol{x})\varphi_{,\alpha\beta}(\boldsymbol{x}) \right) . \qquad (10)$$

The subscript ",$\alpha\beta$" denotes spatial derivatives with respect to the $\alpha$ and $\beta$ components of $\boldsymbol{x}$. In addition to a growing mode $\propto D(t)^2$, the second-order solution has a growing mode $\propto F(t)$, where $F(t)$ is the solution of the equation

$$\ddot{F} + 2H\dot{F} - \frac{3}{2}H^2\Omega F = \frac{3}{2}H^2\Omega D^2. \qquad (11)$$

The ratio $[F(t)/D(t)^2]$ turns out to be nearly independent of time, which implies that it is nearly independent of $\Omega$ (Bouchet et al. 1992),

$$\frac{F}{D^2} \approx \frac{3}{7}\Omega^{-2/63}. \qquad (12)$$

The corresponding second-order term of $\theta$ is then obtained by substituting $\delta_1 + \delta_2$ and $\boldsymbol{v}_1$ into the continuity equation and equating the second order terms,

$$\theta_2(\boldsymbol{x},t) = -f(\Omega)D^2\left\{C\epsilon^2(\boldsymbol{x}) + \boldsymbol{\nabla}\epsilon(\boldsymbol{x})\cdot\boldsymbol{\nabla}\varphi(\boldsymbol{x}) + [1-C]\varphi_{,\alpha\beta}(\boldsymbol{x})\varphi_{,\alpha\beta}(\boldsymbol{x})\right\}, \qquad (13)$$

$$C(\Omega) \equiv \frac{\dot{F}}{2D\dot{D}}. \qquad (14)$$

Much like $F/D^2$ in eq. (12), we find in a similar way that $C$ is also almost independent of $\Omega$,

$$C(\Omega) \approx \frac{3}{7}\Omega^{-1/21}. \qquad (15)$$

This approximation is accurate to within 2% in the range $0.1 \leq \Omega \leq 10$ (Figure 1). A better, empirical approximation, accurate to within 0.5% in the range $0.1 \leq \Omega \leq 1$, is given by $C(\Omega) \approx 3/7\,\Omega^{-0.055}$.

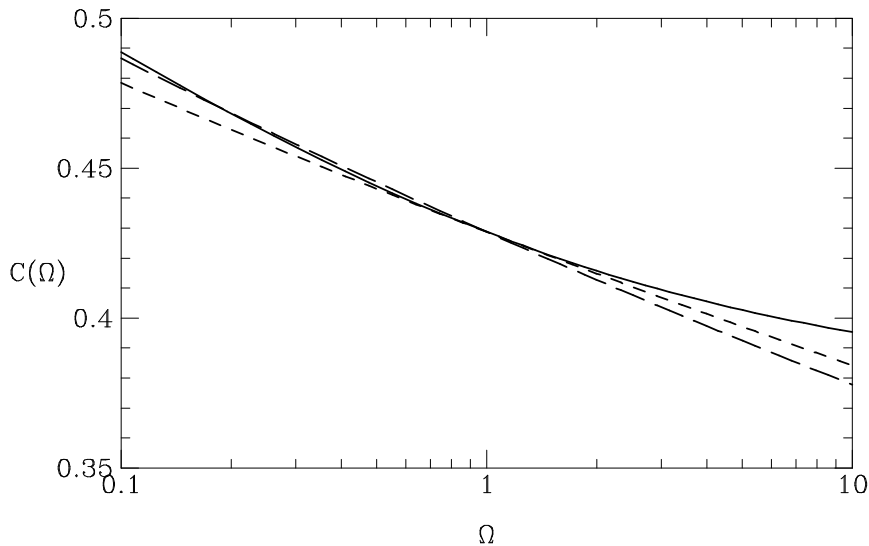

**Fig 1.** The function $C(\Omega)$ of Eq. (14) (solid) compared to the approximate forms $3/7\,\Omega^{-1/21}$ (dashed), and $3/7\,\Omega^{-0.055}$ (long dashed).

Substituting equation (13) in the second term of equation (9) we obtain (*e.g.* following Peebles 1980, §18, but for a general $\Omega$)

$$\langle \theta^3 \rangle = -f(\Omega)[4C + 2] \langle (\delta_1)^2 \rangle^2 , \tag{16}$$

and then, with equation (8),

$$T_3 \equiv \frac{\langle \theta^3 \rangle}{\langle \theta^2 \rangle^2} = -\frac{1}{f(\Omega)}[4C + 2] \approx -\frac{26}{7} \frac{1}{\Omega^{0.6}} . \tag{17}$$

This is a key result, valid at the lowest order in $\langle \theta^2 \rangle$. As expected, the $\Omega$ dependence of $T_3$ is indeed quite strong, which promises that $T_3$ can provide a sensitive measure of $\Omega$.

## 2.2. The Effects of Smoothing

The ratio in equation (17) refers to the statistics of the unsmoothed $\theta(\boldsymbol{x})$ at a point. In practice, however, the observed velocity field must involve significant smoothing of fluctuations at the small-scale end, in order to overcome inevitable noise and non-linear effects. The smoothing is expected to affect $T_3$, as it affects $S_3$.

The velocity field smoothed by a window function $W(\boldsymbol{x})$ is defined by a convolution in real space,

$$\boldsymbol{v}_s(\boldsymbol{x}) = \int d^3\boldsymbol{x}' \, \boldsymbol{v}(\boldsymbol{x}') \, W(\boldsymbol{x} - \boldsymbol{x}') . \tag{18}$$

The divergence of the smoothed velocity, $\boldsymbol{\nabla} \cdot \boldsymbol{v}_s$, equals the smoothed divergence of the velocity, $H\theta_s$, because the convolution and the divergence commute as linear operators. It can therefore be written as the Fourier integral

$$\theta_s(\boldsymbol{x}) = (2\pi)^{-3/2} \int d^3\boldsymbol{k} \, \tilde{W}(\boldsymbol{k}) \, \tilde{\theta}_{\boldsymbol{k}} e^{i\boldsymbol{k}\cdot\boldsymbol{x}} , \tag{19}$$

where $\tilde{W}$ and $\tilde{\theta}$ are the Fourier transforms of $W$ and $\theta$ respectively. The *variance*, to leading order, is then

$$\langle \theta_s^2 \rangle = D^2 (2\pi)^{-3} \int d^3\boldsymbol{k} \, P(k) \, \tilde{W}(\boldsymbol{k})^2 . \tag{20}$$

We used the first order relation $\tilde{\theta}(\boldsymbol{k}) = D\tilde{\epsilon}(\boldsymbol{k})$, where $\tilde{\epsilon}(\boldsymbol{k})$ is the Fourier transform of $\epsilon(\boldsymbol{x})$, and the power spectrum is

$$\langle \tilde{\epsilon}(\boldsymbol{k})\tilde{\epsilon}(\boldsymbol{k}') \rangle = \delta_{dirac}(\boldsymbol{k} + \boldsymbol{k}')P(k) . \tag{21}$$

The *skewness* of the smoothed divergence becomes

$$\langle \theta_s^3 \rangle = -f(\Omega) \, D^4 \, (2\pi)^{-6} \int d^3\boldsymbol{k} \, d^3\boldsymbol{k}' \, P(k) \, P(k') \, \tilde{W}(k) \, \tilde{W}(k') \, \tilde{W}(|\boldsymbol{k}+\boldsymbol{k}'|) \, T(\boldsymbol{k},\boldsymbol{k}') , \tag{22}$$

$$T(\boldsymbol{k}, \boldsymbol{k}') = 6C + 6\mu(k/k') + 6[1 - C]\mu^2 , \qquad (23)$$

with $\mu \equiv (\boldsymbol{k} \cdot \boldsymbol{k}')/kk'$. For a power-law spectrum, $P(k) \propto k^n$, an integral similar to the one above was reduced to an analytic expression by Juszkiewicz et al. (1993a) both for a top-hat and for a Gaussian window, and it is straightforward to rederive $T_3$ using their methods. Recently the calculation for a Gaussian window function was extended beyond the integer values of $n$, considered by Juszkiewicz et al. (1993a), onto arbitrary $n \geq -3$: Łokas et al. (1994), expressed the integral (22) in terms of the hypergeometric function. The resulting skewness parameter is given by

$$T_3 = -\frac{1}{f(\Omega)} \left[ 3\mathrm{F}\left(\frac{n+3}{2}, \frac{n+3}{2}; \frac{3}{2}; \frac{1}{4}\right) + (4C - n - 4)\mathrm{F}\left(\frac{n+3}{2}, \frac{n+3}{2}; \frac{5}{2}; \frac{1}{4}\right) \right] . \qquad (24)$$

The hypergeometric function F should not be confused with the perturbation growth rate $F(t)$. The results for integer $n$ are listed in Table 1; Figure 2 shows $T_3$ as a function of $\Omega$ and $n$. We notice that the dependence of $T_3$ on the type of the window becomes stronger for higher values of $n$, where the small-scale structure is more important. The $\Omega$ dependence remains strong for all spectral slopes.

**Table 1.** Skewness vs. spectral slope for Gaussian and top-hat smoothing windows. $C(\Omega)$ can be read off Figure 1, or using the approximations discussed in the text (eq. 15). The terms in curly brackets have to be added in a future case of galaxy-weighted smoothing, where $b$ is the linear biasing factor of the sampled tracers.

| $n$ | $-f(\Omega)T_3$, Gaussian window function | $-f(\Omega)T_3$, top-hat |
|---|---|---|
| $-3$ | $2 + 4C(\Omega)$ | $2 + 4C(\Omega)$ |
| $-2$ | $6\left(\frac{\pi}{3} - \frac{\sqrt{3}}{2}\right) + 6C(\Omega)\left(\sqrt{3} - \frac{\pi}{3}\right) - \left\{6b\left(1 - \frac{\sqrt{3}}{2}\right)\right\}$ | $1 + 4C(\Omega) - \{b\}$ |
| $-1$ | $6\left(\frac{10\pi}{3\sqrt{3}} - 6\right) + 6C(\Omega)\left(8 - \frac{4\pi}{\sqrt{3}}\right) - \left\{6b\left(\frac{2\pi}{3\sqrt{3}} - 1\right)\right\}$ | $4C(\Omega) - \{2b\}$ |
| $0$ | $6\left(\frac{8\pi}{3} - \frac{44\sqrt{3}}{9}\right) + 6C(\Omega)\left(\frac{16\sqrt{3}}{3} - \frac{8\pi}{3}\right) - \left\{6b\left(1 - \frac{4}{3\sqrt{3}}\right)\right\}$ | $-1 + 4C(\Omega) - \{3b\}$ |
| $1$ | $6\left(\frac{4\pi}{3\sqrt{3}} - \frac{8}{3}\right) + 6C(\Omega)\left(\frac{8}{3} - \frac{8\pi}{9\sqrt{3}}\right) - \left\{6b\left(1 - \frac{4\pi}{9\sqrt{3}}\right)\right\}$ | $-2 + 4C(\Omega) - \{4b\}$ |

Bernardeau (1993) showed that for the *top-hat window*, the 'pure power-law' results of Juszkiewicz et al. (1993a) remain valid when $n$ is not an integer and is varying with the smoothing radius, $R$:

$$T_3 \equiv \frac{\langle \theta_s^3 \rangle}{\langle \theta_s \rangle^2} = -\frac{1}{f(\Omega)}(4C + 2 - \gamma) , \quad \gamma \equiv -\frac{\mathrm{d}\log\langle \theta_s^2 \rangle}{\mathrm{d}\log R} . \qquad (25)$$

Here the spectral index is allowed to vary in the range $0 \geq \gamma < 4$; outside of this range $T_3$ diverges. In the top-hat case, $\theta_s$ represents a local fluctuation in the mean expansion (because it is the integral of $\nabla \cdot v$ in a sphere of radius $R$, which, by the Gauss theorem, is proportional to the total flux of matter across its spherical boundary and thus represents a local perturbation in the Hubble flow). For a pure *power-law* spectrum the variance of $\delta$ in a top-hat sphere of radius $R$ is $\langle \delta^2 \rangle \propto R^{-(3+n)}$, so $\gamma = 3 + n$, and for a top-hat window we have

$$T_3 \approx -\frac{1}{\Omega^{0.6}}\left(\frac{26}{7} - (n+3)\right) \text{, valid for } -3 \leq n < 1, \quad (26)$$

to a good approximation, based on ignoring the $\Omega$-dependence of $C(\Omega)$ and setting $C \approx 3/7$.

Since the Zel'dovich (1970) approximation is frequently used to study weakly-nonlinear dynamics and statistics, it is instructive to rederive $T_3$ from a perturbative expansion of the Zel'dovich solution. We will use the superscript 'Z', to distinguish the result from $T_3$. The Zel'dovich solution gives $\delta_2$, similar to that in eq. (10), but with $F = 0$ (Grinstein & Wise 1987, Bouchet et al. 1992). Therefore, $C = 0$, and the term 26/7 in equation (26) is replaced by 2, giving

$$T_3^Z = -\Omega^{-0.6}(n+1) \neq T_3 . \quad (27)$$

Numerical simulations, discussed in the next section as well as those of Juszkiewicz et al. (1993b) agree with the real perturbation theory ($T_3$) but not with $T_3^Z$. Note, that for spectral slopes $-1 < n < 5/7$, the Zel'dovich approximation fails to reproduce even the sign of $T_3$. This failure is not surprising since the perturbative solutions of the real equations of motion conserve mass as well as momentum to all orders by definition, while the Zel'dovich solution conserves momentum only to first order.

We have also computed the correction for *galaxy*-weighted smoothing, which may have a potentially interesting future application. In this case the distribution of $\theta$ is heavily weighted by densely-sampled regions. Since the sampling density of velocity tracers is correlated with the underlying galaxy density, which is probably correlated with the density of matter ("biasing") and thus with $\theta$, the distribution of $\theta$ smoothed is severely biased. In the ideal case where the density fluctuation of sampled galaxies is proportional to the mass-density fluctuation, $\delta_{sample} = b_{sample}\delta$, we find the smoothing correction to consist of an additional term in eq. (22), $+6bf^{-1}\langle \theta_s^2 \rangle^2$, and an additional term in eq. (23), $+6b(1+ k \cdot k'/k'^2)$. For a top-hat window, the final expression replacing eq. (25) is

$$T_3 = -\frac{1}{f(\Omega)}[4C + 2 - (1 + b_{sample})\gamma] . \quad (28)$$

Future data may have uniform, complete sampling, with $b_{sample}$ the same as $b$, the linear biasing factor of all galaxies. Since the galaxy-weighted $T_3$ depends on $\Omega^{0.6}$ and on $\beta \equiv b/\Omega^{0.6}$ in different ways, it may be used to determine $\Omega$ and $b$ separately. Methods based on comparing the velocity data to the galaxy distribution are limited to determining the degenerate combination $\beta$, so the galaxy-weighted $T_3$ may eventually become an interesting tool. In Table 1, we list in curly brackets the extra terms, which should appear when galaxy-weighted smoothing is applied; the $b$ parameter there denotes $b_{sample}$. For a Gaussian window, the expression in square brackets in eq. (24) should contain the following additional terms: $+b\left\{3\mathrm{F}(\frac{n+3}{2},\frac{n+3}{2};\frac{3}{2};\frac{1}{4})-n\mathrm{F}(\frac{n+3}{2},\frac{n+3}{2};\frac{5}{2};\frac{1}{4})-6\right\}$ . The results obtained in this way are significantly different from those we have just obtained for a volume-weighted smoothing.

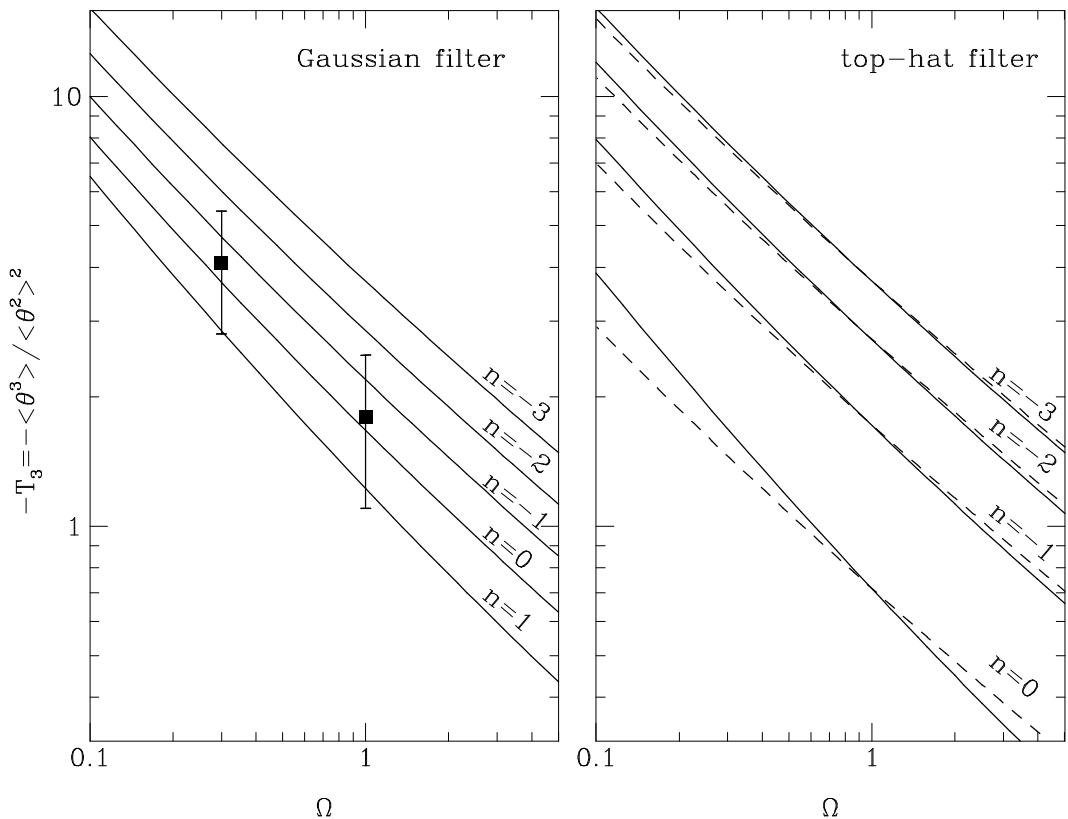

**Fig 2.** The exact second-order values of $-T_3 = -\langle\theta_R^3\rangle/\langle\theta_R^2\rangle^2$ as a function of $\Omega$ for different values of the spectral index $n$ of a power-law spectrum ($n = -3, -2, -1, 0$ etc. from top to bottom). Left: smoothing with a Gaussian window. Right: smoothing with a top-hat window. The filled squares correspond to CDM numerical simulations with $\Omega = 0.3$ and $\Omega = 1.0$, Gaussian smoothed with radius 12 h$^{-1}$Mpc (where $n \approx 0$). The error bars represent the cosmic scatter due the finite-size sample of radius 40 h$^{-1}$Mpc. The dashed lines correspond to approximation (26).

## 3. PRELIMINARY APPLICATION TO SIMULATIONS AND REAL DATA

*3.1. Simulations*

The validity of the second-order approximation for $T_3$ in the case $\Omega = 1$ was confirmed using N-body simulations (Juszkiewicz et al. 1993b). In this paper we use similar simulations for two purposes: to test the $\Omega$ dependence of the second-order approximation derived for $T_3$, and to estimate the cosmic scatter due to the fact that, in practice, $T_3$ is computed from one finite realization of a random field. This is only a preliminary test using simulations; the results are expected to be model dependent, sensitive, in particular, to the shape of the fluctuation power spectrum.

Twelve simulations were run (Nusser & Dekel 1993) for each of the cases, $\Omega = 1$ and 0.3, using a particle-mesh code (Bertschinger and Gelb 1991) with $64^3$ grid cells and particles in a periodic cubic box of comoving size 160 h$^{-1}$Mpc. The initial conditions for each simulation were a random Gaussian realization of the "standard" CDM spectrum (Davis et al. 1985, h= 0.75). The simulations were stopped at a time step when, based on linear growth, the rms density fluctuation in top-hat spheres of radius 8 h$^{-1}$Mpc was unity ($\sigma_8 = 1$). The velocity field was evaluated in a $64^3$ cubic grid of spacing 2.5 h$^{-1}$Mpc by linear cloud-in-cell assignment of velocity to a grid point from the particles in its neighboring cells. The fields were then smoothed on a much larger scale, volume weighted, using a spherical Gaussian window of radius 12 h$^{-1}$Mpc. This procedure ensures smoothing over nonlinear and orbit-mixing regions and it produces quasilinear fields which are comparable in rms amplitude to the fields deduced from observations.

We select from the simulations many independent spheres of radius $R = 40$ h$^{-1}$Mpc, compute $T_3$ within each sphere by summing over the grid points, and obtain the mean and standard deviation of $T_3$ over the spheres. The mean is found to be $T_3 = -1.8$, $-4.1$ for $\Omega = 1$, 0.3 respectively, in pleasant agreement with our second-order predictions (Fig. 2). The cosmic standard deviation at $R = 40$ h$^{-1}$Mpc, for the assumed power-spectrum, is $\epsilon_1 = \pm 0.7$, $\pm 1.3$ respectively. It becomes smaller at larger volumes, of course, but $R \sim 40$ h$^{-1}$Mpc is the practical current limit for high-quality data.

*3.2. Data*

In the present method-oriented paper, we bring only a preliminary application to real data, and defer a careful, detailed application to a paper in preparation. The $\theta$ field is provided by the current POTENT analysis (Dekel et al. 1994; based on Mark III data, Willick et al. 1994). The raw data consist of $\sim 3000$ galaxies with measured redshifts $z_i$ and redshift-independent estimated distances $r_i$ of assumed random Gaussian errors $\sigma_i \simeq (0.15 - 0.21) r_i$. The peculiar velocities are $u_i = z_i - r_i$. The distances are corrected for inhomogeneous Malmquist bias by heavy grouping (Willick 1991) and by adopting a correction procedure which assumes that IRAS 1.2 Jy galaxies trace the underlying galaxy distribution. The radial peculiar velocities are first smoothed with a Gaussian window of radius 12 h$^{-1}$Mpc onto a spherical grid, yielding a radial velocity field $u(\boldsymbol{x})$. The smoothing suppresses the random uncertainties due to distance errors and sparse, nonuniform sampling. The adopted weighting scheme attempts to mimic equal-volume weighting and thus to minimize the associated systematic biases (Dekel, Bertschinger & Faber 1990, DBF). POTENT then recovers the transverse components of the velocity field by requiring that it is derived from a scalar potential, $v = -\nabla\phi$, as predicted by linear gravitational instability and maintained later via Kelvin's circulation theorem under appropriate smoothing (Bertschinger & Dekel 1989). The potential at each point $x$ is computed by integrating $\phi(x) = -\int_0^r u(r', \vartheta, \varphi) dr'$, and the transverse velocity components are derived by differentiating this potential. The divergence field, $\theta$, is computed by finite differencing using a cubic grid of 5 $h^{-1}$Mpc spacing.

The resulting divergence field is contaminated by random errors, $\epsilon_2$, due to the uncertainty in the distance measurements, $\sigma_i$. They are estimated at each grid point by Monte Carlo noise simulations (DBF). The errors at $r = 40$ $h^{-1}$Mpc are typically smaller than $\epsilon_2 = 0.2$. These random errors should not have a systematic effect on the mean and the skewness of $\theta$, but they do add systematically to the true variance. We therefore compute the variance of the Monte-Carlo errors within the 40 $h^{-1}$Mpc sphere and subtract it from the total variance computed for the field recovered by POTENT. The same set of Monte-Carlo noise simulations allow us to estimate the distance error in $T_3$.

The value of $T_3$ within 40 $h^{-1}$Mpc is tentatively found to be $T_3 = -1.1 \pm 0.7$. For the total error, the quoted distance error, $\epsilon_2 = \pm 0.7$, should roughly add in quadrature to the cosmic scatter estimated by the simulations, $\epsilon_1$. Hence, $\Omega = 1$ is fully consistent with the data, while $\Omega = 0.3$ is marginally rejected at the ~2-sigma level.

## 4. DISCUSSION

We sketched a method for measuring $\Omega$ from the probability distribution function (PDF) of the smoothed velocity divergence field under the assumption of mildly-nonlinear gravitational growth from Gaussian initial fluctuations. Unlike the PDF of $\delta$, whose shape in the quasi-linear regime is uniquely determined by its variance, the PDF of $\theta$ at a given variance depends on the dynamics which made it grow, via $\Omega$. In particular, the ratio of moments $T_3 = \langle\theta^3\rangle/\langle\theta^2\rangle^2$ is expected to be $\propto \Omega^{-0.6}$. The expected value of $T_3$ was computed using second-order perturbative approximation and confirmed by N-body simulations. The result takes into account the inevitable effect of smoothing, which depends on an assumed power spectrum.

When applied to the volume-weighted velocity field recovered by POTENT from the observed radial peculiar velocities of galaxies, the method is *free* of assumptions regarding the biasing relation between galaxy density and mass density. The sampled galaxies are assumed to trace the large-scale *velocity* field, like any other test body subject to the large-scale gravitational field and sharing the same initial conditions, but they are *not* assumed to trace the underlying mass distribution. This method is thus ascribed to a unique family of methods which give constraints on $\Omega$ directly (including, *e.g.*, Nusser & Dekel 1993; Dekel & Rees 1994), unlike most methods which measure the degenerate ratio $\Omega^{0.6}/b$ (see Dekel 1994 for a review).

The method relies on the hypothesis that the initial fluctuations were a realization of a *Gaussian* random field. As mentioned in the Introduction, this hypothesis finds support in the observed IRAS galaxy density, but the hypothesis can in fact be tested using the velocity data itself. The scaling relation between the skewness and the square of the variance, $\langle \theta^3 \rangle \propto \langle \theta^2 \rangle^2$, is unique to Gaussian initial conditions. This relation can in principle be measured in the data by varying the smoothing length in the velocity analysis. A practical limitation arises from the limited extent of the available data compared to the sparse sampling, which permits only a limited dynamical range for the smoothing length.

In view of the data available, the method proposed is subject to two kinds of severe *uncertainties*. First, the random errors and biases in the recovered velocity field, due to the distance errors and the sparse nonuniform sampling, which we partly estimate using the POTENT Monte-Carlo noise simulations. Second, the fact that the data consists of only a few tens of independent points and therefore cannot be assumed a "fair sample" and cannot trace the PDF in great detail. The associated cosmic scatter is estimated using N-body realizations of the CDM spectrum, and similar estimates for other possible spectra would be useful. The two errors added in quadrature indicate that the method as applied to the existing data is capable of rejecting an $\Omega = 0.3$ cosmology with marginal confidence. With more data in a more extended volume the discriminatory power of the method will increase.

The $\Omega$ dependence of the PDF of $\theta$ is not limited to the skewness, and $T_3$ is not necessarily the most useful measure of $\Omega$. For example, Bernardeau (1993) computed a similar $\Omega$-dependent function of moments involving the kurtosis of $\theta$ (for top-hat smoothing). In practice, however, the skewness and kurtosis are both sensitive to the poorly-sampled, noisy tails of the distribution. Methods which make use of the whole shape of the PDF, that can be derived analytically in the quasi-linear regime (Bernardeau 1994), might turn out to be more practical.

Our preliminary results favor a high value of $\Omega$, of the order of unity. This result is in accord with the result of Nusser & Dekel (1993), also based on the assumption of Gaussian initial fluctuations, combined with an analysis of the evolution of the PDF different from the analysis adopted here, and a different approximation for gravity in the quasi-linear regime – the Zel'dovich approximation. High $\Omega$ values are also obtained from the velocity divergence near voids (Dekel & Rees 1994), and by comparing the velocity data to the galaxy density (Dekel 1994 for a review). On the other hand, other estimates of $\Omega$, such as those based on the anisotropy of galaxy clustering in redshift space (*e.g.* Cole et al. 1994, and references therein) or variational principle analysis of trajectories of groups of galaxies (Peebles 1993), yield low $\Omega$.

It should be emphasized, that our results obtained from the data provide an illustration of the analysis rather than a definitive measurement of $\Omega$ because the present sets of data are too small. Our control over the sampling errors should improve in the next several years, when deeper surveys become available.

**Acknowledgments**

We thank Adi Nusser for the N-body simulations used here, borrowed from Nusser & Dekel (1993).